\documentclass{aa}
               \usepackage{graphicx,natbib,psfig}
\newcommand{\ltsima}    {$\;    \buildrel    <   \over    \sim    \;$}
\newcommand{\gtsima}    {$\;    \buildrel    >   \over    \sim    \;$}
\newcommand{\lta}     {\lower.5ex\hbox{\ltsima}}     \newcommand{\gta}
{\lower.5ex\hbox{\gtsima}}

\begin{document}  

\title{The  Chandra  view of  the  3C/FR~I  sample  of low  luminosity
radio-galaxies}
  
\titlerunning{The Chandra view of FR~I radio-galaxies}
  
\author{Barbara Balmaverde  \inst{1} \and Alessandro  Capetti \inst{2}
\and Paola Grandi \inst{3}}

\offprints{B. Balmaverde}
     
\institute{Dipartimento  di Fisica  Generale, Universit\'a  di Torino,
Via      P.Giuria      1,      I-10125      Torino,      Italy      \\
\email{balmaverde@ph.unito.it} \and INAF - Osservatorio Astronomico di
Torino,  via   Osservatorio  20,  I-10025  Pino   Torinese,  Italy  \\
\email{capetti@to.astro.it}  \and  INAF   -  Istituto  di  Astrofisica
Spaziale e Fisica Cosmica, Via  Gobetti 101, I-40129 Bologna, Italy \\
\email{grandi@bo.iasf.cnr.it}}

\date{}
   
\abstract{We present results from  Chandra observations of the 3C/FR~I
sample  of low  luminosity  radio-galaxies.  We  detected a  power-law
nuclear component in 12 objects out of the 18 with available data.  In
4 galaxies  we detected  nuclear X-ray absorption  at a level  of $N_H
\sim (0.2-6)  \times 10^{22}$  cm$^{-2}$.  X-ray absorbed  sources are
associated with the  presence of highly inclined dusty  disks (or dust
filaments projected  onto the  nuclei) seen in  the HST  images.  This
suggests  the existence  of a  flattened X-ray  absorber, but  of much
lower optical depth than in classical obscuring tori.  We thus have an
un-obstructed view toward most FR~I nuclei while absorption plays only
a marginal role in the remaining objects.

Three pieces  of evidence support  an interpretation for a  jet origin
for the  X-ray cores: i)  the presence of strong  correlations between
the  nuclear  luminosities in  the  radio,  optical  and X-ray  bands,
extending  over  4  orders  of  magnitude  and  with  a  much  smaller
dispersion ($\sim$ 0.3 dex) when  compared to similar trends found for
other classes of AGNs, pointing to a common origin for the emission in
the three bands;  ii) the close similarity of  the broad-band spectral
indices with the sub-class of BL Lac objects sharing the same range of
extended  radio-luminosity, in  accord with  the  FR~I/BL~Lacs unified
model; iii) the presence of  a common luminosity evolution of spectral
indices in both FR~I and BL~Lacs.

The low luminosities of the  X-ray nuclei, regardless of their origin,
strengthens  the interpretation  of  low efficiency  accretion in  low
luminosity radio-galaxies.

 } \maketitle
 
\section{Introduction}  

By exploring the properties of low luminosity radio-galaxies (LLRG) it
is possible to  extend the study of active  galactic nuclei (AGN), and
in  particular of  the  radio-loud sub-population,  toward the  lowest
level of  nuclear luminosity. This  offers the opportunity  to improve
our  understanding of  the mechanism  of accretion  onto super-massive
black holes  and of their radiative manifestations.   In this respect,
it  is important  to note  that LLRG  not only  represent the  bulk of
radio-loud active  galaxies, due to the steepness  of their luminosity
function,  but also  a substantial  fraction of  the  overall galaxies
population. In fact LLRG are associated to a fraction as high as 40 \%
of     all    bright     elliptical     and    lenticular     galaxies
\citep{sadler89,auriemma77}.   Thus  the   presence  of  a  low  power
radio-galaxy  represents  by  far  the most  common  manifestation  of
nuclear activity in early type galaxies.

Low  luminosity AGN  (LLAGN) also  represent a  link between  the high
luminosity AGN and the population  of quiescent galaxies, as it is now
widely recognized that most (if not all) galaxies host a super-massive
black-hole   \citep[e.g.][]{kormendy95}.    The   comparison  of   the
different  manifestations  of  nuclear  activity  across  the  largest
possible  range of luminosity  is then  a crucial  step to  unveil the
connections  (and  the  diversities)  between  active  and  non-active
galaxies.

Unfortunately, the  study of LLRG  has been significantly  hampered by
the contamination  from host galaxy emission which,  in most observing
bands, dominates the emission from the AGN.  To constrain the physical
processes  at  work  in  these  objects it  is  clearly  necessary  to
disentangle   the   AGN   and   host's  contributions   via   spectral
decomposition  or high  resolution imaging.   In the  last  few years,
thanks  in particular  to the  Hubble Space  Telescope (HST)  this has
become routinely possible.

An  example  of the  insights  that  can  be obtained  following  this
approach  comes  from  the   results  derived  from  the  analysis  of
broad-band HST imaging of  a sample of LLRG \citep{chiaberge:ccc}.  In
the  majority of  the targets,  HST  images revealed  the presence  of
unresolved optical  nuclei.  Fluxes and luminosities  of these sources
show a  tight correlation  with the radio  cores, extending  over four
orders  of magnitude.  This  has been  interpreted as  being due  to a
common  non-thermal emission process  in the  radio and  optical band,
i.e.   that we  are seeing  the optical  emission from  the base  of a
relativistic jet.  The low luminosity  of LLRG optical nuclei, and the
possible dominance  of the  emission related to  out-flowing material,
indicates a low  level of accretion and/or a  low radiative efficiency
with  respect to  classical,  more luminous  AGN.   The high  fraction
($\sim$  85  \%) of  objects  with  detected  optical nuclear  sources
suggests  a  general  lack  of  obscuring molecular  tori,  a  further
distinction with respect  to other classes of AGN.   The tenuous torus
structure, when  considered together with the low  accretion rate, the
low mass of  the compact emission line regions  (10 - 10$^3 M_{\sun}$)
and  the limits  to the  mass  of the  Broad Line  Region ($M_{BLR}  <
10^{-2} M_{\sun}$),  indicates that  a general paucity  of gas  in the
innermost  regions   of  LLRG  emerges  as   the  main  characterizing
difference from more powerful AGN \citep{capetti:cccriga}.

The  advent of  Chandra provides  us  with the  unique opportunity  to
extend the  study of  LLRG to  the X-ray band  since its  high spatial
resolution  enables  us  to  isolate  any  low  power  nuclear  source
associated to  LLRG.  Furthermore, with  respect to the  optical data,
the spectral  capabilities of Chandra  allow us to study  directly the
spectral behaviour of  the X-ray nuclei. This can  be used to quantify
the effects  of local absorption  as well as  the slope of  their high
energy emission as to better  constrain the emission processes at work
in  these sources,  particularly when  combined in  a multi-wavelength
analysis using also optical and radio observations.  This approach was
already    followed     by    several    authors     in    the    past
\citep[e.g.][]{fr1sed,hardcastle00,trussoni03,hardcastle03}         but
clearly, the  superior capabilities  of Chandra warrant  to re-explore
this issue in much greater depth.

Here we present results obtained from the analysis of Chandra data for
the  sample of  LLRG  drawn  from the  3C  catalogue of  radio-sources
\citep{bennett62},  more  specifically   those  with  a  morphological
classification as FR~I  \citep{fanaroff74}.  Considering LLRG from the
3C  sample, the  same studied  by \citet{chiaberge:ccc}  represents an
obvious choice. It  is the best studied sample  of radio-loud galaxies
in existence with a vast suite of ground and spaced based observations
for  comparison  at  essentially  all accessible  wavelengths.   Other
samples  of LLRG  have been  considered  in the  literature, but,  not
surprisingly, the coverage provided  by Chandra observations for these
samples is  severely incomplete. For example,  only about 10  \% of B2
sample of LLRG  \citep{colla75} has been observed with  Chandra (to be
compared to the $\sim$ 55 \% for the 3C/FR~I sample), and the analysis
is further  obstructed by the  unknown selection biases  introduced in
the choice of the individual targets.  Similarly we preferred to focus
on the  Chandra data alone, to  provide the highest  possible level of
uniformity in the analysis.

The paper  is organized as follows:  in Sec. \ref{obs}  we present the
observations and the data reduction that lead to the results described
in Sec.   \ref{results} and  in particular to  the detection  of X-ray
nuclei in most FR~I.  In Sec.  \ref{absorption} we explore the effects
of the nuclear  X-ray absorption, by also relating  it to the presence
of  dust features  seen in  the HST  images. The  origin of  the X-ray
nuclei found in  LLRG is explored in Sec.  \ref{jet}.  The results are
discussed and summarized in Sec. \ref{summary}.

We adopted $\rm{H}_o=75 $ km s$^{-1} $Mpc$^{-1}$ and $q_0=0.5$.
 
\section{Chandra data analysis}  
\label{obs}

We selected the same  sample considered by \cite{chiaberge:ccc} formed
by   the  33   radio-sources  from   the  3CR   catalogue   with  FR~I
morphology\footnote{The only addition is 3C~189, which is an FR~I part
of the  3C sample  \citep{edge59} but not  of its revised  version 3CR
\citep{bennett62}.}.

We searched  for Chandra observations available in  the public archive
up  to January  2005 and  we found  data for  18 objects  (see Table\,
\ref{observations}).  All  observations we considered  were made using
the Advanced CCD Imaging  Spectrometer (ACIS-I and ACIS-S) without any
transmission  grating in place.   When more  than one  observation was
available, we generally choose the one with the longer exposure times;
only for  3C~338 two observations  of approximately equal  length were
present  and  they  were  fitted simultaneously.   For  the  brightest
objects instead we choose the  observation with the lowest time frame,
to reduce the pile-up effect.

\begin{table} 
\caption{Log of the observation.}
\begin{tabular}{l c c c c c } \hline
  Name   &  Obs Id  & Inst & Exp time & Frame time \\
         &          &      &   [ks]   &  [s]         \\
\hline
3C 028   & 3233 & ACIS/I & 50.38 & 3.1\\
3C 031   & 2147 & ACIS/S & 44.98 & 3.2\\
3C 066B  & 828  & ACIS/S & 45.17 & 1.5\\
3C 075   & 4181 & ACIS/I & 21.78 & 3.2\\
3C 078   & 3128 & ACIS/S & 5.23  & 3.1\\
3C 083.1 & 3237 & ACIS/S & 95.14 & 3.1\\
3C 084   & 3404 & ACIS/S & 5.86  & 0.4\\
3C 189   & 858  & ACIS/S & 8.26  & 0.8\\
3C 270   & 834  & ACIS/S & 35.18 & 1.8\\
3C 272.1 & 803  & ACIS/S & 28.85 & 3.2\\
3C 274   & 1808 & ACIS/S & 14.17 & 0.4\\
3C 296   & 3968 & ACIS/S & 50.08 & 3.2\\
3C 317   & 890  & ACIS/S & 37.23 & 3.2\\
3C 338   & 497  & ACIS/S & 19.72 & 3.2\\
         & 498  & ACIS/S & 19.16 & 3.2\\
3C 346   & 3129 & ACIS/S & 46.69 & 0.8\\ 
3C 348   & 1625 & ACIS/S & 15.00 & 3.2\\
3C 438   & 3967 & ACIS/S & 47.9  & 3.1\\
3C 449   & 4057 & ACIS/S & 29.56 & 3.2 \\
\hline
\end{tabular}
\label{observations}
\medskip
\end{table}

We reduced all  the data using the Chandra  data analysis CIAO v3.0.2,
with  the  CALDB version  2.25.   Therefore,  to  correct for  the  QE
degradation we create  a new ARF (effective area  files) using the new
file 'acisD1999-08-13contamN0003.fits' released with the CALDB version
2.26 (as recommended  by the ``ACIS Modeling and  Analysis Team").  We
reprocessed all the data from level  1 to level 2, using the CIAO tool
acis\_process\_event  with  all  parameters  restored to  the  default
values, i.e. applying the pixels and PHA randomization.  The new event
file were filtered selecting a standard set of ASCA grades (02346) and
rejecting time interval of high background levels.

Our aim is the study of the X-ray emission associated to the AGN.  For
this  reason we  choose as  extracting region  a circle  of $1\farcs2$
radius  centered on  the  emission  peak (for  the  fainter nuclei  we
checked their position using HST or VLA observation) and as background
region    a    source-centered     annulus    with    radii    between
$1\farcs2-2\farcs5$.   When X-ray  emission  from an  extended jet  is
present,  we excluded  from  the background  region the  corresponding
angular portion.   With this  strategy we aim  at reducing as  much as
possible  the diffuse  thermal  X-ray component,  as  to maximize  the
nuclear contribution.

A possible drawback of this approach, i.e.  a small extraction region,
is  related  to the  variations  of  the  Point Spread  Function  with
energy. These might  cause the inclusion of a  varying fraction of the
nuclear emission in our aperture  in the different bands.  However, we
tested  that  the  PSF  effects  are negligible.   In  particular  the
fraction of nuclear counts falling in the background annulus decreases
only from $\sim$ 8 \% in the 0.5-2  keV band to $\sim$ 2 \% in the 2-5
keV band.  This has a  marginal impact on the derived luminosities and
spectral indices.

The  spectra were  modeled using  XSPEC version  11.3.0. We  used only
events with energy $>0.5$ keV, where the calibration is best known and
the background  contribution is negligible.  We  rebinned the spectrum
to a minimum of 25 counts per bin in order to ensure the applicability
of  the  $\chi^2$ statistics.   We  modeled  all  the spectra  with  a
superposition   of   a   thermal   and   a   power-law   (non-thermal)
component. For  the thermal component the metal  abundances were fixed
to 0.5 \citep{fabbiano89}  and the absorption was set  to the galactic
value; for the power-law component we allowed for the presence of {\sl
local} absorption.   The relative complexity of the  model prevents us
from  obtaining a  robust  estimate  of all  free  parameters for  the
sources with  a low count rate.  More  quantitatively, we experimented
that  only  for the  twelve  objects for  which  the  spectrum can  be
regrouped  in at  least 15  channels, as  to leave  us with  $\gta$ 10
degree of freedom, corresponding to approximately $\sim$ 400 total net
counts, the fit is sufficiently constrained.  In the six sources below
this threshold we set upper  limits for the non-thermal fluxes fitting
their spectra using only a  power-law model (with column density fixed
at the galactic  level).  We initially fixed the power  law index to 2
but, to  be conservative, we then look  for the value of  Gamma in the
range 1.5 - 2.5 corresponding to the largest flux value.

This is a  quite conservative approach but our  choice it is supported
by the fact that in these objects we cannot accurately disentangle the
contribution  of  the  thermal  and  non-thermal  components;  a  less
cautious analysis  is prone to  lead to highly uncertain  estimates in
particular of the AGN X-ray flux.  For the 3 brightest sources (3C~78,
3C~84 and  3C~274) we included the  correction for the  effects of the
pile-up  applying  the  algorithm  proposed by  \citet{davis01}.   The
results of the fit are given in Table\, \ref{tab_lett2}.

For the  12 objects  in which  we performed a  two components  fit, we
tested with an F-test\footnote {Protassov et al. (2002) suggested that
it might  inappropriate to  use the F-test  to test an  added spectral
component. However, because at the moment there is not known any other
simple alternative method, we decided in any case to apply the F-test,
above  all  considering  that  the  presence of  an  extended  thermal
emission and a point-like nuclear source are also directly attested by
the  image.}  that  the  presence  of both  a  thermal  component  and
non-thermal  always corresponded  to an  improvement of  the fit  at a
level $>$ 95\%.

Errors  and  upper  limits  are  quoted at  90\%  confidence  for  one
parameter of  interest ($\Delta  \chi^2$=2.71), with the  exception of
$N^z_H$ and $\Gamma$, well known  to be strongly coupled, for which we
used a 90\%  confidence level for two parameters  of interest ($\Delta
\chi^2$=4.61).

\citet{donato04}  considered X-ray  observations  of a  sample of  low
luminosity  radio-galaxies,  largely   overlapping  with  the  objects
discussed here. With the exception of 3C~338, 3C~348 and 3C~449, which
we more  conservatively defined as  upper limits, there is  a complete
agreement between  the two lists  of detected X-ray nuclei.   Also the
temperature of the thermal component as well as the power-law spectral
index  are in very  good accord  when they  are measured  from Chandra
data.   Considering  the values  of  $N^z_H$,  since  we treated  more
conservatively the  errors on this quantity, we  obtained upper limits
in a few objects which Donato  et al. considered as detection of local
absorption.

\begin{table*} 
\caption{
Results  of the  fit: (1)  source name,  (2) Galactic  hydrogen column
density  [10$^{22}$ cm$^{-2}$], (3) local  hydrogen column
density  [10$^{22}$ cm$^{-2}$], (4)  power-law photon  index $\Gamma$,
(5) temperature of  the thermal component [keV], (6)  observed flux  of the
thermal  component corrected  for Galactic  absorption  [erg cm$^{-2}$
  s$^{-1}$], (7)  non-thermal nuclear flux  (corrected for absorption)
at 1  keV [erg cm$^{-2}$  s$^{-1}$], (8) $\chi^2$/degrees  of freedom.
When the number of d.o.f. was smaller than 10, no fit was performed. *
marks  sources affected  by pile-up  and corrected  with  the Davies's
algorithm.}

\hspace{0.1cm} 
\begin{tabular}{l r r r r r r r r r} \hline

Name     & N$_{\rm H,gal}$ & N$_H^z$    & $\Gamma$            & KT                  & F$_{x,ther}$(0.5-5 keV)   & F$_{x,nuc}$ (1 keV)       & $\chi^2/dof$  \\
\hline
3C 028       & 0.05   &  --                 &   --                &         --          &          --               & $<1.5$ E-14               & --/0       \\
3C 031       & 0.05   & $<0.3$              & $1.4_{-0.4}^{+0.3}$ & $0.6_{-0.1}^{+0.1}$ & $3.3_{-0.7}^{+0.8}$ E-14  & $1.9_{-0.5}^{+1.0}$ E-14  & $29.4/24$  \\
3C 066B      & 0.09   & $<0.7$              & $2.4_{-0.4}^{+0.7}$ & $0.3_{-0.1}^{+0.1}$ & $6.2_{-2.7}^{+1.4}$ E-14  & $1.2_{-0.4}^{+0.8}$ E-13  & $30.3/37$  \\
3C 075       & 0.09   & --                  &    --               &      --             &       --                  &  $<3.6$ E-14              & --/2         \\
3C 078$^*$   & 0.10   & $<0.4$              & $2.2_{-0.3}^{+0.4}$ & $0.2_{-0.1}^{+0.1}$ & $1.6_{-1.5}^{+0.8}$ E-13  & $7.4_{-4.1}^{+4.6}$ E-13  & $39.4/27$  \\
3C 083.1     & 0.15   & $2.5_{-1.4}^{+0.8}$ & $2.2_{-0.5}^{+0.9}$ & $0.5_{-0.1}^{+0.1}$ & $2.7_{-0.3}^{+0.7}$ E-14  & $6.3_{-4.1}^{15.2}$ E-14  & 7.19/19    \\
3C 084$^*$   & 0.15   & $<0.1$              & $1.6_{-0.1}^{+0.1}$ & $0.3_{-0.1}^{+0.1}$ & $7.6_{-1.6}^{+3.5}$ E-13  & $2.8_{-0.1}^{+0.5}$ E-12  & $150.1/161$ \\
3C 189       & 0.05   & $<0.2$              & $2.1_{-0.4}^{+0.6}$ & $0.3_{-0.1}^{+0.2}$ & $4.5_{-3.4}^{+5.4}$ E-14  & $1.3_{-0.2}^{+0.5}$ E-13  & $22.7/16$   \\
3C 270       & 0.01   & $5.7_{-3.4}^{+3.2}$ & $0.8_{-0.4}^{+0.7}$ & $0.5_{-0.1}^{+0.1}$ & $6.6_{-0.6}^{+0.7}$ E-14  & $8.4_{-5.4}^{13.3}$ E-14  & 54.15/50    \\
3C 272.1       & 0.02   & $0.2_{-0.1}^{+0.3}$ & $2.1_{-0.3}^{+0.2}$ & $0.5_{-0.2}^{+0.3}$ & $1.0_{-0.7}^{+1.3}$ E-14  & $5.1_{-1.8}^{+3.7}$ E-14  & $17.9/23$   \\
3C 274 $^*$  & 0.02   & $<0.03$             & $2.3_{-0.1}^{+0.1}$ & $0.7_{-0.1}^{+0.1}$ & $1.2_{-0.4}^{+0.4}$ E-13  & $7.2_{-0.5}^{+0.9}$ E-13  &$118.6/98$   \\
3C 296       & 0.02   & $1.0_{-0.6}^{0.6}$  & $1.6_{-0.4}^{+0.5}$ & $0.5_{-0.1}^{+0.1}$ & $3.2_{-0.4}^{+0.3}$ E-14  & $6.5_{-2.0}^{+3.4}$ E-14  & $27.81/33$  \\
3C 317       & 0.03   & $<0.1$              & $2.0_{-0.2}^{+0.2}$ & $0.2_{-0.1}^{+0.2}$ & $1.1_{-0.9}^{+0.8}$ E-14  & $6.3_{-0.7}^{+0.9}$ E-14  &33.6/38      \\
3C 338       & 0.01   & --                  &  --                 & --                  &--                         & $<3.3$ E-14               & --/8        \\
3C 346       & 0.05   & $<0.1$              & $1.7_{-0.1}^{+0.1}$ & $0.6_{-0.3}^{+0.6}$ & $1.5_{-1.1}^{+1.1}$ E-14  & $1.5_{-0.1}^{+0.1}$ E-13  &92.1/89      \\
3C 348       & 0.06   & --                  &  --                 & --                  &--                         & $<3.1$ E-14               & --/0          \\
3C 438       & 0.21   & --                  &  --                 & --                  &--                         & $<1.5$ E-14               & --/0          \\
3C 449       & 0.11   & --                  &  --                 & --                  &--                         & $<2.1$ E-14               & -- /0         \\
\hline
\end{tabular}
\label{tab_lett2}
\end{table*}

 \begin{table*} 
\caption{Summary of radio, optical and X-ray data of the sample: (1) source name, (2) redshift of the galaxy, (3) nuclear radio flux at 5 GHz [mJy],
(4) nuclear radio luminosity at 5 GHz [erg s$^{-1}$], (5) extended radio luminosity at 178 MHz [erg s$^{-1}$], (6) optical nuclear luminosity at 5500
\AA\ [erg s$^{-1}$],
(7) unabsorbed X-ray nuclear flux (0.5-5 keV) [erg cm$^{-2}$ s$^{-1}$], (8) X-ray nuclear luminosity (0.5-5 keV) corrected for absorption [erg s$^{-1}$].
Note: Information about redshift, radio core flux, extended radio emission, optical luminosities are from 
\citet{chiaberge:ccc}. We adopt an optical spectral index $\alpha_o = 1$.}
\begin{tabular}{l r r r r r r r r r} \hline

Name        &   Redshift & $F_c$ (5 GHz)& $\nu L_c$ (5 GHz) &   $\nu L_t$ (178 MHz)    & $\nu L_o$   & $F_x$ (0.5-5 keV) & $L_x$ (0.5-5 keV ) \\
				
\hline
		
3C~028      & 0.1952     &  $<$0.2     & $<$7.3 E38& 2.1 E42 &$<$1.3 E41 & $<3.4$ E-14               & $<2.5$ E42                \\
3C~031      & 0.0169     &    92.0     & 2.5 E39   & 1.6 E40 & 4.5 E40   & $6.2_{-1.7}^{+3.1} $ E-14 & $3.4_{-0.1}^{+0.2} $ E40  \\ 
3C~066B     & 0.0215     &   182.0     & 8.0 E39   & 3.8 E40 & 2.4 E41   & $2.4_{-0.8}^{+1.6} $ E-13 & $2.1_{-0.1}^{+0.1} $ E41  \\
3C~075      & 0.0232     &    39.0     & 2.0 E39   & 1.9 E40 &  --       & $<8.3$ E-14               & $<8.5$ E40	         \\
3C~078      & 0.0288     &   964.0     & 7.6 E40   & 5.5 E40 & 2.1 E42   &  $8.8_{-3.6}^{+1.5}$ E-13 &  $1.4_{-0.1}^{+0.1}$ E42  \\ 
3C~083.1    & 0.0251     &    21.0     & 1.3 E39   & 5.7 E40 & 9.3 E39   & $1.3_{-0.9}^{+3.1}$ E-13  &  $1.6_{-0.1}^{0.3}$ E41   \\
3C~084      & 0.0176     &   42370.0   & 1.2 E42   & 4.2 e40 & 4.9 E42   & $6.1_{-0.1}^{+0.8} $ E-12 & $3.6_{-0.1}^{+0.1} $ E42  \\ 
3C~189      & 0.043      &    195      & 3.4 E40   & 7.1 E40 & 6.2 E41   & $2.8_{-0.6}^{+1.1} $ E-13 & $9.9_{-0.1}^{+0.1} $ E41  \\
3C~270      & 0.0074     &   308.0     & 1.6 E39   & 1.0 E40 & 2.9 E39   & $4.4_{-2.9}^{+6.9}$ E-13  & $4.6_{-0.3}^{+0.8}$ E40  \\
3C~272.1    & 0.0037     &   180.0     & 2.3 E38   & 9.0 E38 & 8.5 E39   & $1.5_{-0.4}^{+0.8} $ E-13 & $3.9_{-0.1}^{+0.3} $ E39   \\
3C~274      & 0.0037     &  4000.0     & 5.2 E39   & 4.9 E40 & 5.6 E40   & $1.3_{-0.1}^{+0.2} $ E-12 & $3.4_{-0.1}^{+0.1} $ E40  \\
3C~296      & 0.0237     &    77.0     & 4.1 E39   & 2.5 E40 & 2.0 E40   & $1.8_{-0.5}^{+1.0} $ E-13 & $1.9_{-0.1}^{+0.1} $ E41  \\
3C~317      & 0.0342     &   391.0     & 4.4 E40   & 1.9 E41 & 1.2 E41   & $1.5_{-0.2}^{+0.2} $ E-13 & $3.3_{-0.1}^{+0.1} $ E41	 \\
3C~338      & 0.0303     &   105.0     & 9.2 E39   & 1.4 E41 & 9.7 E40   & $<7.6$ E-14               & $<1.3$ E41               \\
3C~346      & 0.1620     &   220.0     & 5.5 E41   & 9.7 E41 & 6.4 E42   & $4.0_{-0.3}^{+0.2} $ E-13 & $2.0_{-0.1}^{+0.1} $ E43 \\
3C~348      & 0.1540     &    10.      & 2.3 E40   & 2.8 E43 & 2.0 E41   & $<7.3$ E-14               & $<3.3$ E42               \\
3C~438      & 0.2900     &    17.0     & 1.4 E41   & 1.3 E43 &$<$3.5 E41 & $<3.4$ E-14               & $<5.4$ E42                \\
3C~449      & 0.0181     &    37.0     & 1.1 E39   & 1.3 E40 & 6.2 E40   & $<4.8$ E-14               & $<3.0$ E40	         \\
		  
\hline

\end{tabular}
\label{tab_lett}

\medskip
\end{table*}

\section{Results}
\label{results}

Before  we proceed to  discuss our  results, it  is important  to note
that, since archival  data are available for a  fraction of about half
of the  original sample  of 3C/FR~I,  it is possible  that we  are not
considering  an  unbiased sub-sample.  We  therefore compared  various
parameters (e.g. redshift, core and total radio-power) for the 3C/FR~I
sample   as   a   whole   and   for  our   sub-sample.    Applying   a
Kolmogorov-Smirnov  test   we  found  that   their  distributions  are
consistent with being drawn  from the same population with probability
always larger than  60 \%. Only considering the  fraction of optically
nucleated galaxies in the Chandra sub-sample is slightly larger, being
83\%   to   be  compared   to   a  70   \%   for   the  whole   sample
\citep{chiaberge:ccc}. Overall,  it appears that  the galaxies studied
here are  well representative  of the entire  3C/FR~I sample.   On the
other hand,  several galaxies with  somewhat peculiar radio-morphology
are excluded, (e.g. the  fat double 3C~314.1, the compact radio-source
associated to 3C~293, the X-shaped 3C~305, or the complex radio-source
in  3C~433) possibly  leading to  a  higher fraction  of objects  with
classical core-jets FR~I morphology.

\subsection{Local absorption in X-ray nuclei}

\begin{figure} \resizebox{\hsize}{!}
{\includegraphics[width=\textwidth,angle=0]{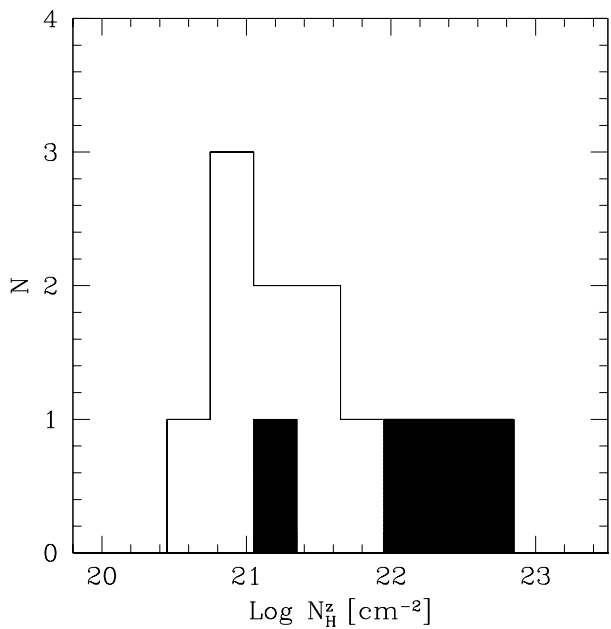}}
\caption{Histogram of  intrinsic absorption in Log  N$_H^z$. The black
 rectangles represent detections, while the white ones represent upper
 limits (at 90\% confidence level).}
\label{nh}
\end{figure}

We obtained a statistically  significant estimate for the local column
density in four objects  (namely 3C~83.1, 3C~270, 3C~272.1 and 3C~296)
with values  $N^z_H \sim  0.2 - 6  \times 10^{22}$ cm$^{-2}$.   In the
other   cases   (8/12)   we   find   an   upper   limit   to   $N^z_H$
(Table\,\ref{tab_lett2}) ranging from  $0.03 \times 10^{22}$ cm$^{-2}$
to  $0.7 \times 10^{22}$  cm$^{-2}$.  In  Fig.\,\ref{nh} we  present a
histogram  of  the  values  of  the intrinsic  absorption:  the  white
rectangles represent upper  limits, while the black ones  are the four
detections.

\subsection{Relations between radio, optical and X-ray emission.}

\begin{figure*}
\centerline{                 \psfig{figure=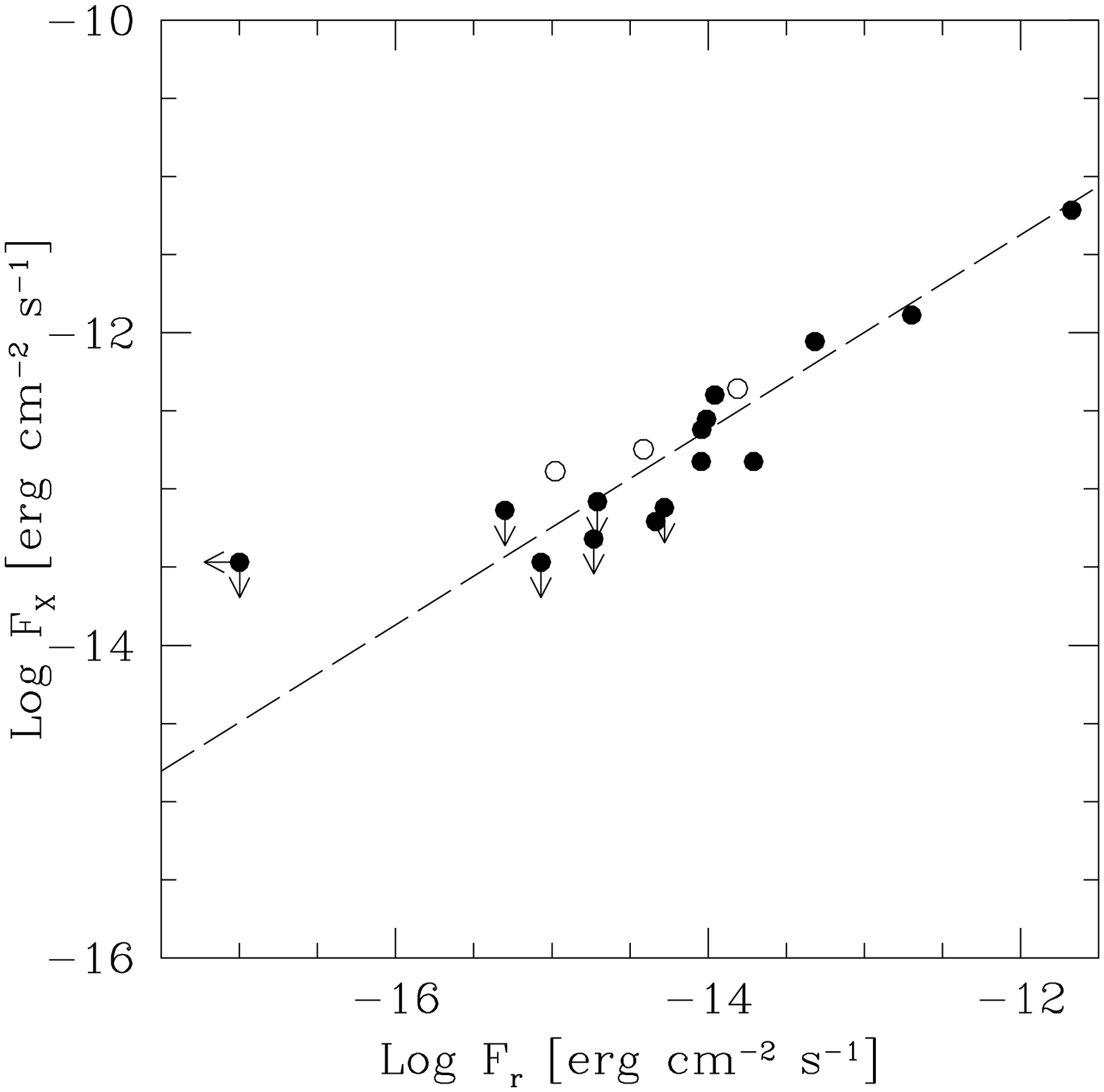,width=0.50\linewidth}
\psfig{figure=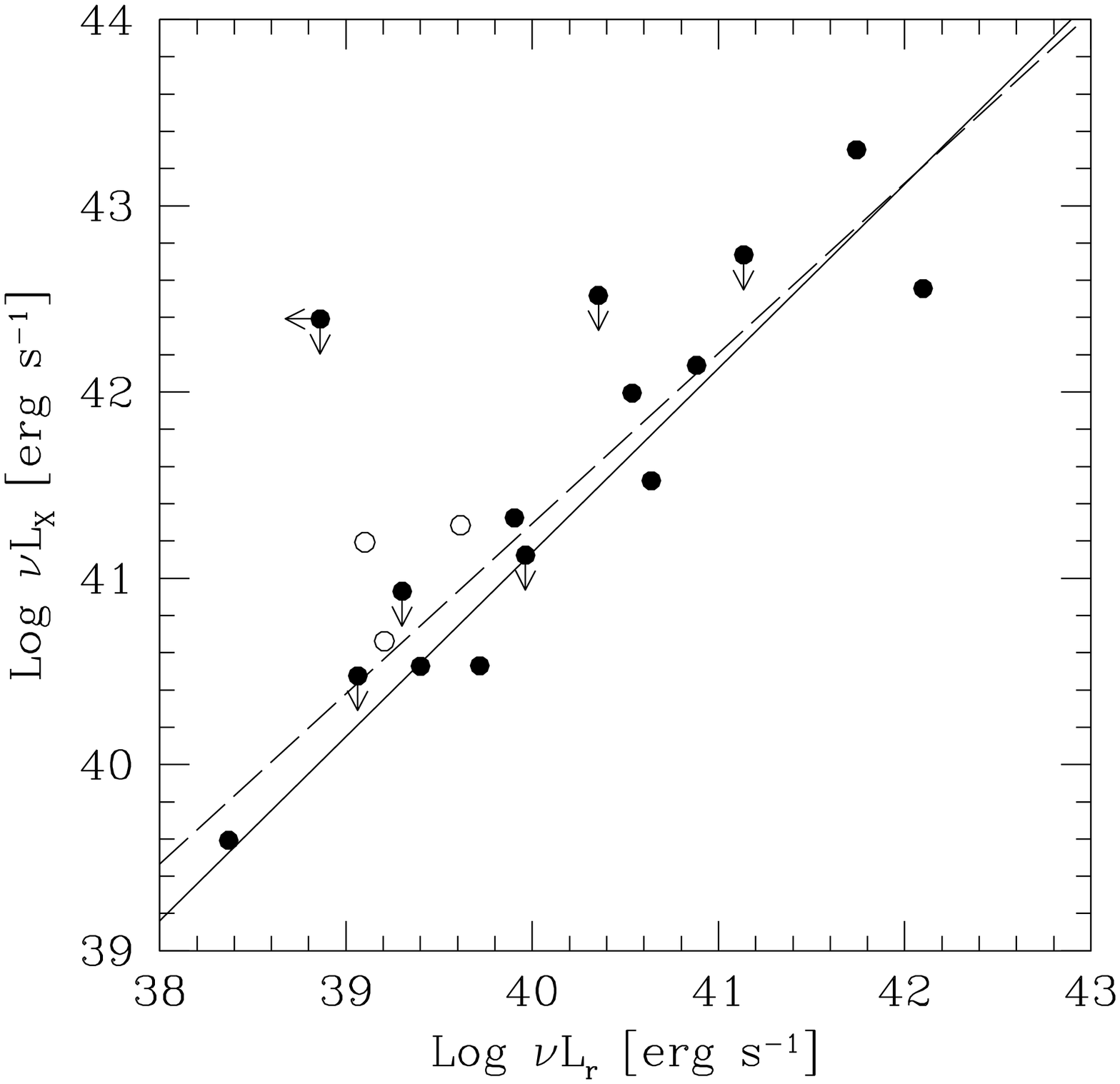,width=0.50\linewidth}}                \centerline{
\psfig{figure=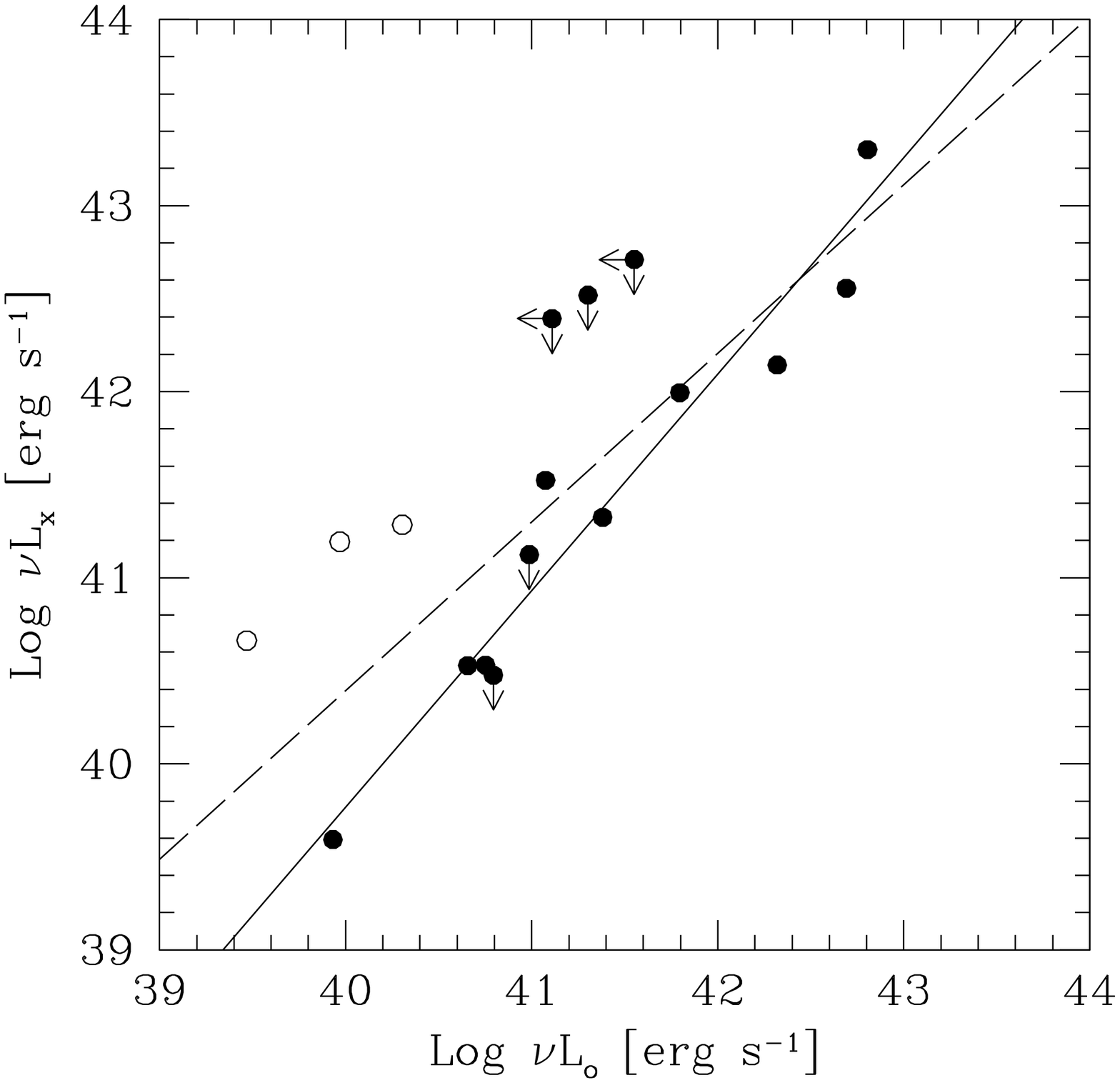,width=0.50\linewidth}
\psfig{figure=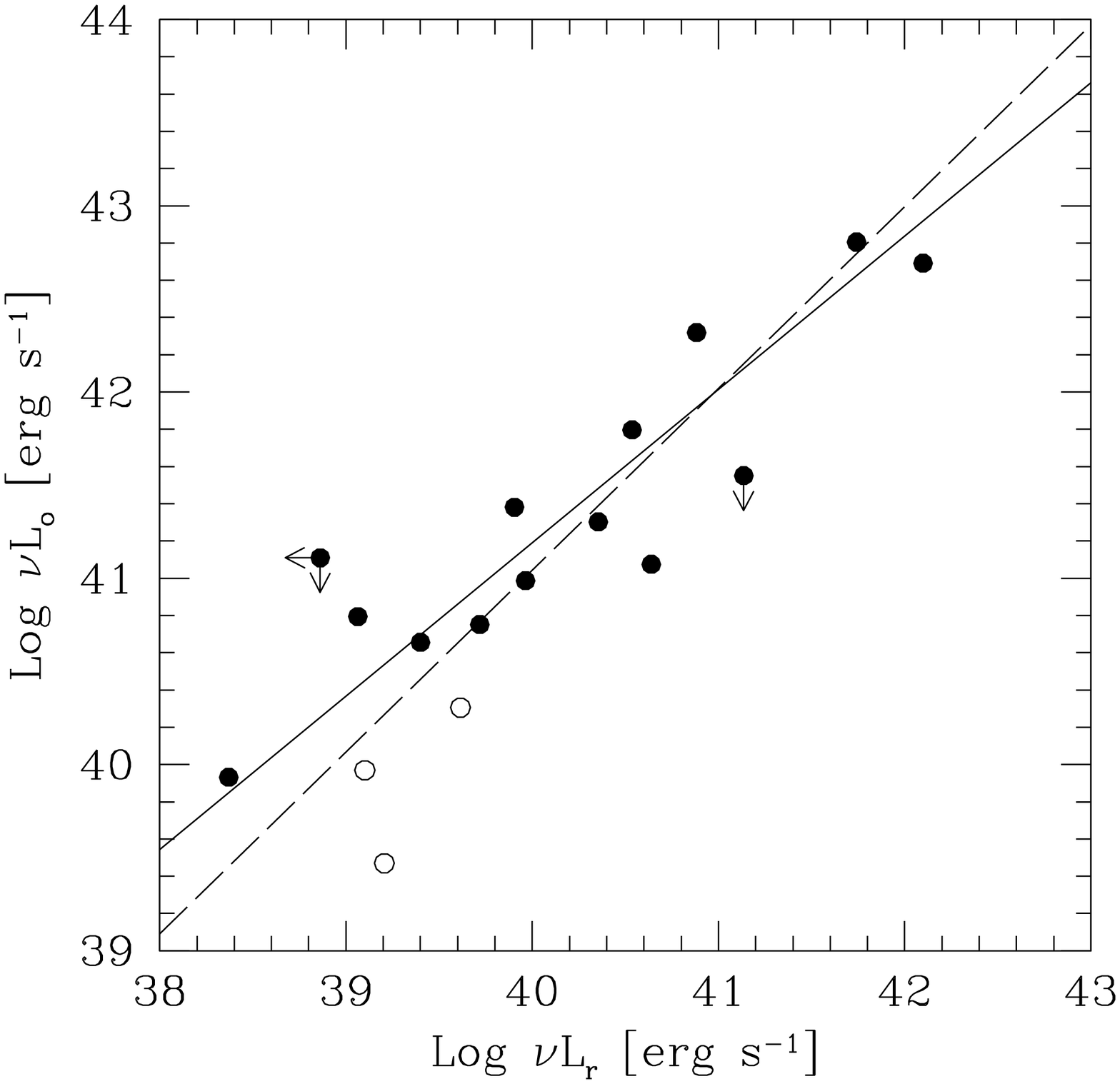,width=0.50\linewidth}}
\caption{Comparison of  (top left)  radio (5 GHz)  and X-ray (0.5  - 5
keV) nuclear fluxes, (top right) radio and X-ray luminosities, (bottom
left)  optical  (5500  \AA)  and X-ray  luminosities,  (bottom  right)
optical  and  radio luminosities.   The  empty  circles represent  the
objects for which we have  detected an intrinsic absorption $> 10^{22}
$cm$^{-2}$.   The solid  line reproduces  the best  linear  fit having
excluded the 3 X-ray absorbed nuclei, while the dashed line is the fit
on the whole sample.}
\label{corr} 
\end{figure*}

In Fig.\, \ref{corr} (top left panel) we compare the nuclear fluxes in
the  radio and  X-ray bands.   A  strong correlation  is present:  the
generalized  (including the  presence of  upper limits)  Spearman rank
correlation coefficient  $\rho$ = 0.88 corresponding  to a probability
that the correlation is not present of P = 0.001.

The best linear  fits were derived as the bisectrix  of the fits using
the two  quantities as independent variables  following the suggestion
by  \citet{isobe90}  that  this  is preferable  for  problems  needing
symmetrical treatment of the  variables.  The presence of upper limits
in the independent variable suggests  to take advantage of the methods
of survival analysis proposed by e.g. \citet{schmitt85}.  However, the
drawbacks  discussed by  \citet{sadler89} and,  in our  specific case,
also  the  non-random distribution  of  upper  limits, disfavour  this
approach.   We therefore preferred  to exclude  upper limits  from the
linear regression analysis. Nonetheless we note that, a posteriori, 1)
the  objects  with  undetected  nuclear  component in  the  X-ray  are
consistent with the correlation defined by the detections only; 2) the
application of  the Schmidt  methods provides correlation  slopes that
agree, within the errors, with our estimates.

A similarly tight correlation between radio and X-ray nuclear emission
is   present   also   considering   luminosities,   with   $\rho$=0.72
corresponding  to  a probability  P=0.003  (see Fig.\,\ref{corr},  top
right panel); it  extends over 4 orders of magnitude,  with a slope of
$m$  = 0.91  $\pm$ 0.12  and  a dispersion  of 0.39  dex (see  Table\,
\ref{tab0} for a summary of  the statistical analysis).  We tested the
possible  influence  of redshift  in  driving  this correlation  (both
quantities  depend  on $z^2$)  estimating  the  Spearman partial  rank
coefficient\footnote{The Spearman Partial Rank correlation coefficient
estimates  the linear  correlation coefficient  between  two variables
taking  into account the  presence of  a third.  If A  and B  are both
related  to the variable  z, the  Spearman correlation  coefficient is
$\rho_{AB,z}=\frac{\rho_{AB}-\rho_{Az}\rho_{Bz}}{[(1-\rho_{Az}^2)(1-\rho_{Bz}^2)]^{1/2}}$}.
The  effect  is negligible  as  the  value  of $\rho$  is  essentially
unchanged.

\begin{table}
\label{corr-sec}
\caption{Correlations summary}
\begin{tabular}{l l c c c c } \hline
Var. A    & Var. B &  $\rho_{AB}$ (P)  & $\rho_{AB,z}$ & Slope & rms     \\
\hline                                          
F$_{x}$ & F$_{r}$     & 0.88 (0.001)  & --       & 0.62$\pm$0.14  & 0.27 \\                                  
L$_{x}$ & L$_{r}$     & 0.72 (0.003)  & 0.67     & 0.91$\pm$0.12  & 0.39  \\
L$_{x}$ & L$_{r}corr$ & 0.79 (0.003)  & 0.77     & 0.99$\pm$0.11  & 0.33  \\
L$_{x}$ & L$_{o}$     & 0.71 (0.004)  & 0.67     & 0.91$\pm$0.16  & 0.52  \\
L$_{x}$ & L$_{o}corr$ & 0.86 (0.002)  & 0.85     & 1.16$\pm$0.09  & 0.23  \\
L$_{o}$ & L$_{r}$     & 0.79 (0.002)  & 0.75     & 0.98$\pm$0.11  & 0.39  \\
L$_{o}$ & L$_{r}corr$ & 0.79 (0.005)  & 0.76     & 0.82$\pm$0.11  & 0.32  \\ 
\hline
\end{tabular}
\label{tab0}

\smallskip
$\rho_{AB}$  is  the  generalized Spearman's  correlation  coefficient
(computed including  upper limits) and in  parenthesys the probability
that there is no  correlation between the variables.  $\rho_{AB,z}$ is
the partial rank coefficient  having excluded the common dependence of
luminosities A and  B from redshift.  The 3  X-ray absorbed nuclei are
excluded in the analysis marked with 'corr'.
\end{table}

We find a similarly strong connection between the X-ray versus optical
nuclear luminosity  (Fig.\,\ref{corr}, left bottom  panel). Again, all
X-ray  upper  limits  are   consistent  with  the  $L_{O}$  vs.  $L_X$
correlation defined by the detected objects. Conversely, note that all
three  sources where we  measured the  largest X-ray  absorbing column
densities (namely 3C~83.1, 3C~270  and 3C~296) lie significantly above
the  correlation between  $L_{O}$ and  $L_X$. This  is  not surprising
since, while  the estimated X-ray fluxes are  de-absorbed, the optical
measurements are  not corrected for the presence  of local absorption.
Indeed, the location of these  X-ray absorbed sources in the $L_{O}$ -
$L_X$  is suggestive  of significant  absorption also  in  the optical
band.  We therefore preferred  to re-estimate the correlation omitting
these three objects, shown in  Fig.\,\ref{corr} with a solid line.  We
obtain a smaller dispersion of 0.23 dex and a steeper corrected slope,
$m$ =1.16$\pm$0.09.

Also  in  the radio-optical  plane  $L_r$ -  $L_O$  we  find a  strong
correlation  as  already  found  by \cite{chiaberge:ccc},  see  Fig.\,
\ref{corr}, bottom  right panel.  Again, the 3  X-ray absorbed sources
are located below  the best fit, supporting the  interpretation for an
optical  deficit derived  from their  location  in the  $L_O$ -  $L_X$
plane.   Similarly,  we  re-estimated  the correlation  excluding  the
potentially  absorbed  objects  in  the  optical.  The  slope  of  the
corrected correlation  is $m =0.82\pm0.11$ (very similar  to the slope
of  the correlation found  by \cite{chiaberge:ccc},  $m =0.88\pm0.12$,
using the whole 3C/FR~I sample) with a dispersion of 0.32 dex.

\begin{figure*}
\centerline{ \psfig{figure=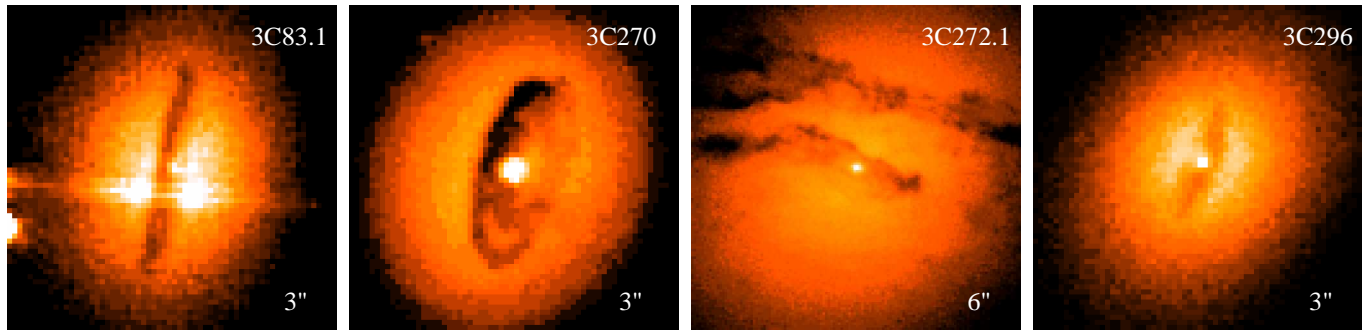,width=\linewidth}}
\caption{HST images  of the  4 radio-galaxies in  which we  detected a
local X-ray absorption, taken from \citet{chiaberge:ccc}.}
\label{hst} 
\end{figure*}

\section{The X-ray absorbing medium in FR~I nuclei}
\label{absorption}

In 4 galaxies  we obtained a detection of  local X-ray absorption.  In
all cases our  estimates of the column density  are in close agreement
with previously published  measurements.\footnote{See Chiaberge et al.
(2002) and Donato et  al. (2004) for 3C~270; Harris  et al. (2002) and
Donato et al. (2004) for 3C272.1; Hardcastle et al. (2005) for 3C~296}
It is particularly instructive the inspection of their HST images (see
Fig.\,\ref{hst}) as they unveil a connection between optical and X-ray
absorption.  The  three sources for  which we found the  largest X-ray
absorption show circumnuclear dusty disk structures, on scales between
300 pc  - 1 kpc.  All disks  are seen at large  inclinations, the less
inclined being associated  to 3C~270 whose axis form  with the line of
sight an  angle $\theta \geq 64^{\circ}$  \citep{jaffe96}.  The fourth
X-ray absorbed galaxy is 3C~272.1:  its HST image show the presence of
a  low optical  depth ($A_V  \sim 0.54$,  \citet{bower97}) filamentary
dust lane, directly projected onto the nucleus.

By no  means dust is  exclusively found in  these 4 sources.   The HST
images of the remaining radio-galaxies also show the presence of dust,
even in the  form of disks. For example 3C~66B  and 3C~78 have face-on
disks \citep{sparks00}  while disks at higher  inclinations (but still
substantially less inclined  than in the 3 cases  discussed above) are
found in  3C~31 and 3C~449.  Off-nuclear dust,  usually distributed in
filaments, is seen also in  3C~84, 3C~274, 3C~317 and 3C~338.  In most
of these  sources, the quality of  the X-ray spectra  is sufficient to
set stringent limits to the column density, $N_H < 10^{22} $cm$^{-2}$.

Therefore  it appears  that X-ray  absorption is  associated  with the
presence of  dust, but only  when i) the  dust intercepts our  line of
sight to  the nucleus or  ii) it takes  the form of a  highly inclined
disk.    While  the  first   case  is   rather  obvious,   the  second
situation\footnote{This behaviour  is shared also  by the radio-galaxy
B2 0055+30,  with a large column  density $N_H ~\sim  7 \times 10^{22}
$cm$^{-2}$ \citep{donato04}  associated to  a highly inclined  disk in
the HST  image \citep{capetti00}.} indicates that  the gas responsible
for  the X-ray  absorption shares  the flattened  distribution  of the
dust, inducing a dependence of  $N_H$ on the viewing angle.  Note that
evidence for a dependence of  absorption on orientation in FR~I nuclei
was  already  found by  \citet{chiaberge:uv}  from  the comparison  of
optical and UV images of radio-galaxies.

However, the column densities toward FR~I nuclei are much smaller than
those of the  optical and X-ray absorbers predicted  (and required) by
the AGN  unified models, i.e.  the circumnuclear obscuring tori;  as a
comparison,  in the  obscured  Seyfert  2 galaxies  75\%  have $N_H  >
10^{23}$ cm$^{-2}$  and about half of  them are Compton  thick ($N_H >
10^{24}$  cm$^{-2}$  \citep{risaliti99}).   Unfortunately,  we  cannot
constrain the location of the  X-ray absorbing gas, other that it must
be coplanar  with the  large scale dust.   It might be  also cospatial
with the dust  and thus being distributed on a  kpc scale, rather than
on the characteristic pc scale of tori in Seyfert galaxies.

Let's now  focus on the three  sources for which we  found the largest
X-ray absorption.   In Sect.\,\ref{corr} we noted  that their location
with  respect  to  the $L_O$  vs  $L_X$  as  well  as $L_O$  vs  $L_r$
correlations can  be interpreted as  due to absorption in  the optical
band \footnote{This interpretation is supported by the extremely steep
UV/optical  nuclear spectrum  of 3C~270  \citep{fr1sed} which  must be
ascribed  to absorption.}.   The optical  absorption can  be estimated
quantitatively  from their offsets  from the  best fit  relations.  We
find $A_R\sim 2 - 4$ from  the $L_O$ vs $L_X$ plane and slightly lower
values, $A_R\sim 1.5 - 3$, from  the $L_O$ vs $L_R$ plane.  The ratios
between optical and X-ray absorption are  in the range $A_V / N_H \sim
(0.7 - 3.3) \times 10^{-22}$ mag cm$^2$, having transformed $A_V$=1.34
$A_R$ using  the \citet{cardelli89} extinction law.   These values are
slightly smaller than  the standard galactic ratio, $A_V  / N_H \sim 5
\times 10^{-22}$ mag  cm$^2$ \citep{bohlin78}. \citet{maiolino01} also
found evidence  for a low  $A_V / N_H$  ratio in AGN, but  with values
extending down to values as low  as $A_V / N_H \sim 3 \times 10^{-24}$
mag cm$^2$. They  interpreted this finding as an  evidence for dust in
the  circumnuclear  region  of  AGNs being  different  from  Galactic,
possibly due  to a dominance  of larger grains.  Clearly,  our results
indicate a much  less extreme departure for the $A_V  / N_H$ value and
consequently from the galactic dust properties.

The fraction  of X-ray absorbed sources  also sets a limit  to the gas
covering factor.  Three out of 12  objects (25\%) are  associated to a
column  density  larger  than  $10^{22}$ cm$^{-2}$.   As  a  reference
values,  in the CfA  sample of  Seyfert galaxies  \citet{huchra92} the
fraction of obscured objects is  53\%, considering only type 2, but it
might as high  as 78\% if also the intermediate types  1.8 and 1.9 are
included  \citet{osterbrock1993}.  Given  the  small (and  incomplete)
sample   statistics  and  the   different  (and   somewhat  arbitrary)
definition  of 'absorbed'  FR~I  or Seyfert,  a  direct comparison  is
probably not appropriate. Certainly however, although X-ray absorption
in  FR~I is  associated to  low  column densities,  it is  not a  rare
phenomenon.

\section{A jet-origin for the X-ray cores?}
\label{jet}
The central issue  for our understanding of the  properties of LLRG is
the origin  of their  nuclear emission.  The  analysis of  the results
presented  in  Section   \ref{corr-sec}  provides  several  pieces  of
evidence, presented in the following three subsections, supporting the
interpretation that the  radio, optical and X-ray cores  have a common
origin, most likely being produced  by the base of a relativistic jet,
as  already suggested  from  the analysis  of  ROSAT and  HST data  by
\citet{worrall94} and \citet{chiaberge:ccc}.

\subsection{Multiwavelength nuclear correlations}
The  first   indication  comes  from   the  presence  of   the  strong
correlations between  the nuclear luminosities in  the three available
bands.

A  connection  between  the  radio,  optical and  X-ray  radiation  is
observed  also in other  classes of  AGNs including  Seyfert galaxies,
radio-galaxies  and QSO  \citep[see  e.g.][]{falcke95}.  However,  the
dispersion   between  the   emission   in  the   different  bands   is
substantially larger  than observed in  our sample. For example,  at a
given radio  luminosity, the X-ray  emission spans over 4-5  orders of
magnitude;  even projecting  this correlation  in a  plane considering
also the  different black hole  masses, the rms  is still as  large as
0.88 dex  \citep{merloni03}.  These results witness, on  the one hand,
the  close link  between accretion  and ejection  in AGN  but,  on the
other, the relatively large  allowed range for the parameters coupling
the  two  processes  and  their  radiative  manifestations,  e.g.  the
strength  of the  magnetic field  and the  structure of  the accretion
disk.

Conversely, the rms of the data  points for our FR~I sample around the
best fit are typically a mere  factor of 2, despite the large range of
luminosities   involved.     Furthermore,   concerning   the   radio-X
correlation,  it  must also  be  noted  that  part of  the  dispersion
certainly arises from the (often substantial) errors associated to (at
least) the  X-ray luminosities,  see Table \,\ref{tab_lett2}.   If one
considers  that  also  other  effects  contribute  to  the  dispersion
(e.g. variability in non simultaneous observations) the rms value must
be considered as a strict upper limit to the intrinsic dispersion.

Thus our  results radically differ  from a quantitative point  of view
from the general trend found in  other AGNs and point to a much closer
connection, such as  a common origin for the  radio, optical and X-ray
emission.

\subsection{Broad band spectral indices}
\label{spix}
\begin{figure} \resizebox{\hsize}{!}
{\includegraphics[width=\textwidth,angle=0]{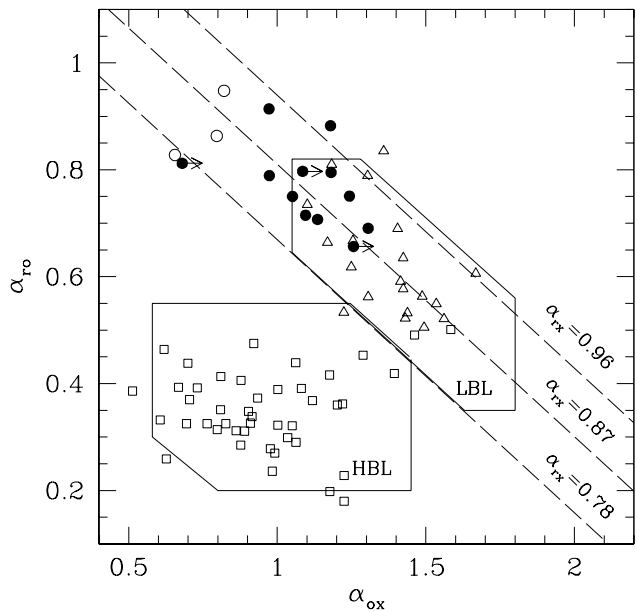}}
\caption{Broad-band spectral indices,  calculated with luminosity at 5
GHz, 5500 \AA\  and 1 keV; the dashed  lines represent constant values
for  the third  index  $\alpha_{rx}$.  The  solid  lines indicate  the
regions of the plane within 2 $\sigma$ from the mean $\alpha_{ro}$ and
$\alpha_{ox}$  for BL  Lacs drawn  from  the Deep  X-Ray Radio  Blazar
Survey (DXRBS) \citet{padovani03}.  Note that FR~I cover approximately
the same portion of the plane of LBL.}
\label{roox}
\end{figure}

The issue  of the  origin of the  X-ray emission  in FR~I can  also be
addressed concentrating on the  Spectral Energy Distributions (SED) of
FR~I compared  to those of other  classes of AGNs. As  a SED behaviour
can  be  parameterized  using  the  broad-band  spectral  indices,  we
combined the available  data in the different energy  bands (scaled to
the standard reference values, i.e. the radio at 5 GHz, the optical at
5500  \AA\  and  X-ray luminosity  at  1  keV)  and we  calculate  the
broad-band  spectral indices  of our  objects. They  are  presented in
Table\,\ref{sindices} and reproduced graphically in the ($\alpha_{ox}$
vs $\alpha_{ro}$)  plane (see Fig.\,\ref{roox}).   The distribution of
spectral indices is in complete agreement with the results obtained by
\citet{hardcastle00}, who used ROSAT data for the X-ray measurements.

All FR~I sources  are confined in the upper left  region of this plane
and  show  a remarkable  homogeneity  in  their  spectral shapes  with
$\alpha_{ox}=1.13\pm    0.11$,    $\alpha_{ro}=0.77\pm   0.08$,    and
$\alpha_{rx}=0.90\pm  0.05$. The  quoted  uncertainties represent  the
measured dispersions of each index, not considering the X-ray absorbed
sources. These  latter objects are slightly offset  from the remaining
of the sample  but, as we argued in Sect.  \ref{corr-sec}, this can be
explained as  due to the  uncorrected dust absorption for  the optical
data.

\begin{table} 
\label{bbspix}
\caption{Broad-band spectral indices.}
\hspace{1.5cm} 
\begin{tabular}{l c c c c } \hline
  Name    &  $\alpha_{ox}$  & $\alpha_{ro}$ &$\alpha_{rx}$ \\
          &                          &            &             &                     \\
\hline      
3C 028    &    --                      &     --             &    --                    \\
3C 031    &   $1.24_{-0.33}^{+0.24}$   &    $0.75\pm0.12$   &   $0.92_{-0.07}^{+0.01}$  \\
3C 066B   &   $1.13_{-0.29}^{+0.58}$   &    $0.71\pm0.01$   &   $0.85_{-0.10}^{+0.20}$   \\
3C 075    &    --                      &      --            &   $>0.83$                  \\
3C 078    &   $1.09_{-0.48}^{+0.54}$   &    $0.71\pm0.01$   &   $0.84_{-0.16}^{+0.19}$   \\
3C 083    &   $0.65_{-0.57}^{+2.10}$   &    $0.83\pm0.01$   &   $0.77_{-0.19}^{+0.72}$   \\
3C  84    &   $1.18_{-0.18}^{+0.23}$   &    $0.88\pm0.09$   &   $0.98_{-0.01}^{+0.05}$   \\
3C 189    &   $1.05_{-0.13}^{+0.33}$   &    $0.75\pm0.01$   &   $0.85_{-0.05}^{+0.11}$   \\
3C 270    &   $0.82_{-0.56}^{+1.38}$   &    $0.95\pm0.01$   &   $0.90_{-0.19}^{+0.47}$   \\
3C 272.1    &   $1.31_{-0.31}^{+0.63}$   &    $0.69\pm0.01$   &   $0.90_{-0.10}^{+0.22}$   \\
3C 274    &   $1.18_{-0.07}^{+0.12}$   &    $0.79\pm0.02$   &   $0.93_{-0.02}^{+0.04}$   \\
3c 296    &   $0.80_{-0.28}^{+0.46}$   &    $0.86\pm0.04$   &   $0.84_{-0.09}^{+0.16}$   \\
3C 317    &   $0.97_{-0.13}^{+0.15}$   &    $0.91\pm0.04$   &   $0.93_{-0.03}^{+0.04}$   \\
3C 338    &   $>1.08$                  &    $0.80\pm0.04$   &   $>0.90$                \\
3C 346    &   $0.97_{-0.13}^{+0.13}$   &    $0.79\pm0.06$   &   $0.85_{-0.02}^{+0.02}$   \\
3C 348    &   $>0.68$                  &    $0.81\pm0.06$   &   $>0.77$                \\
3C 438    &   --                       &    $>0.92$         &   $>0.84$                \\
3C 449    &   $>1.2$                   &    $0.66\pm0.02$   &   $>0.86$                  \\
\hline
\end{tabular}
\label{sindices}
\medskip
\end{table}

We then compare these values  with similar estimates for other classes
of AGNs. Not surprisingly, radio quiet quasars have a median values of
$\alpha_{ro}=-0.1$  and $\alpha_{rx}=0.36$ \citep{elvis94}  much lower
than  our radio  loud sources  and this  applies also  to  the Seyfert
nuclei \citep{ho01}.   But also the  steep spectrum radio  loud quasar
have   significantly   smaller  values   $\alpha_{ro}   =  0.46$   and
$\alpha_{rx} = 0.71$. These differences in spectral indices correspond
to an  optical (and X-ray)  excess for a  given radio core power  by a
factor between 20 and 40.

Conversely,  the FR~I  sources  show a  substantial  overlap with  the
region of the  plane populated by the Low  energy peaked BL~Lacs (LBL,
\citet{padovani95}), see  Fig.\, \ref{roox}.  For  the comparison with
BL~Lacs we consider the radio selected BL Lacs sample derived from the
1Jy catalog \citep{stickel91} and the  BL Lac sample selected from the
{\it  Einstein} Slew  survey \citep{elvis92,perlman96}.   We  used the
classification into  high and low energy  peaked BL Lacs  (HBL and LBL
respectively),  as  well  as   their  multiwavelength  data  given  by
\citet{fossati98}. The  same result is  obtained using BL  Lac objects
drawn   from  the   Deep  X-Ray   Radio  Blazar   Survey   (DXRBS)  by
\citet{padovani03}.

In  BL~Lacs the  emission process  is well  established and  the whole
nuclear radiation is ascribed to  the base of a relativistic jet.  The
similarity in  the spectral indices  distributions of FR~I  and BL~Lac
provides further  support to  the idea that  also in FR~I  the nuclear
emission is non-thermal emission associated to an outflow.

\begin{figure*}
\centerline{                 \psfig{figure=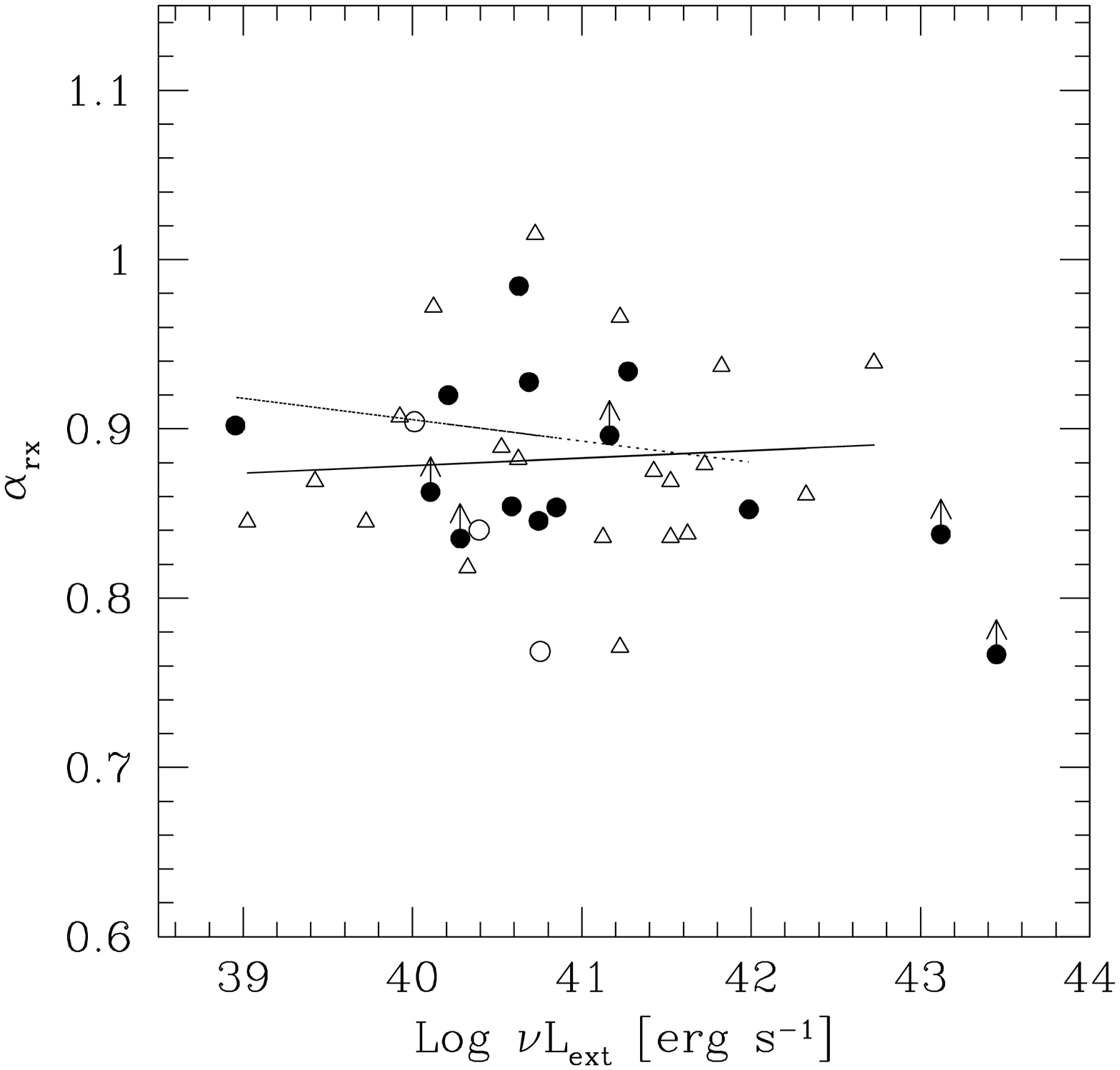,width=0.33\linewidth}
\psfig{figure=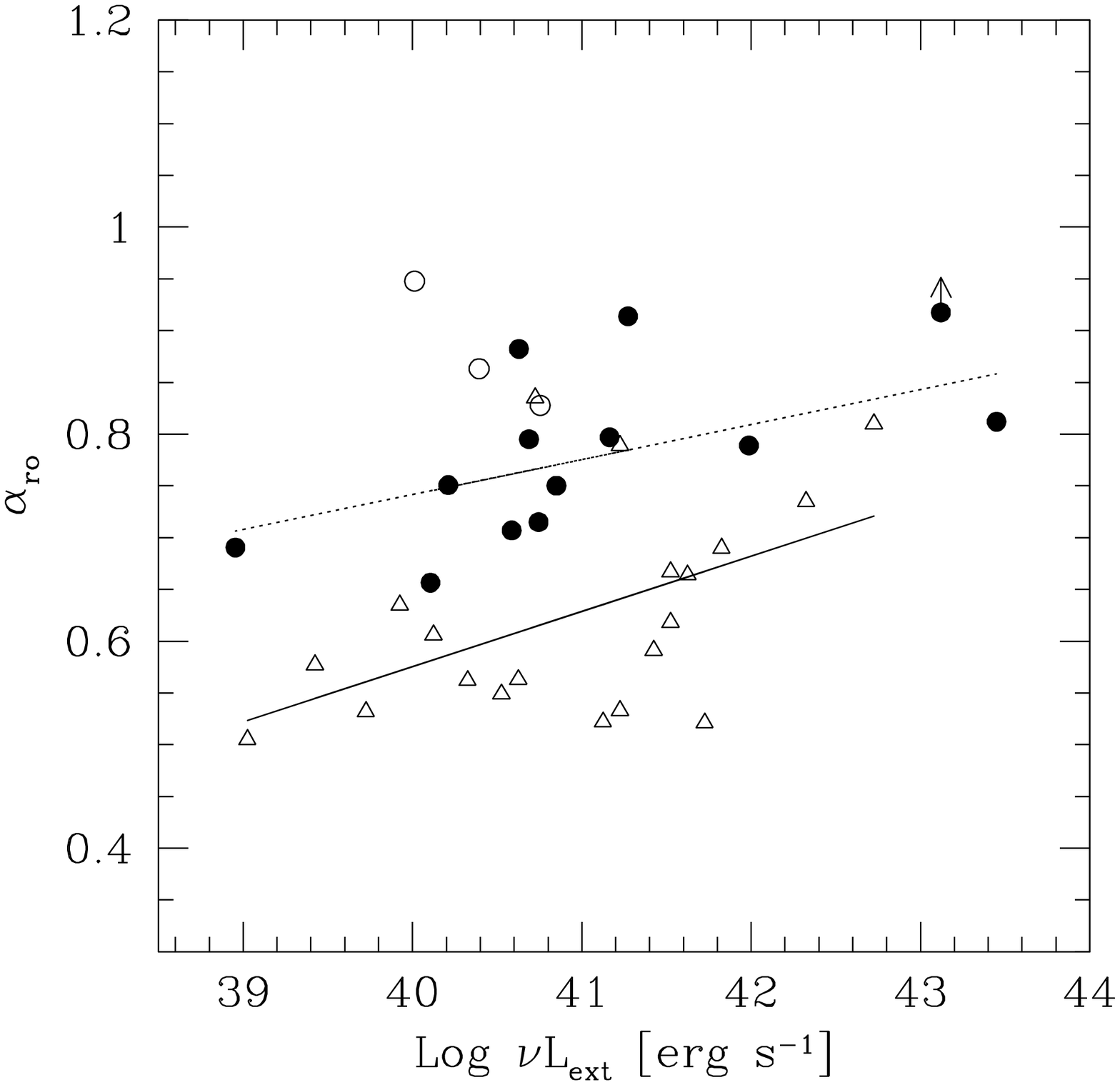,width=0.33\linewidth}
\psfig{figure=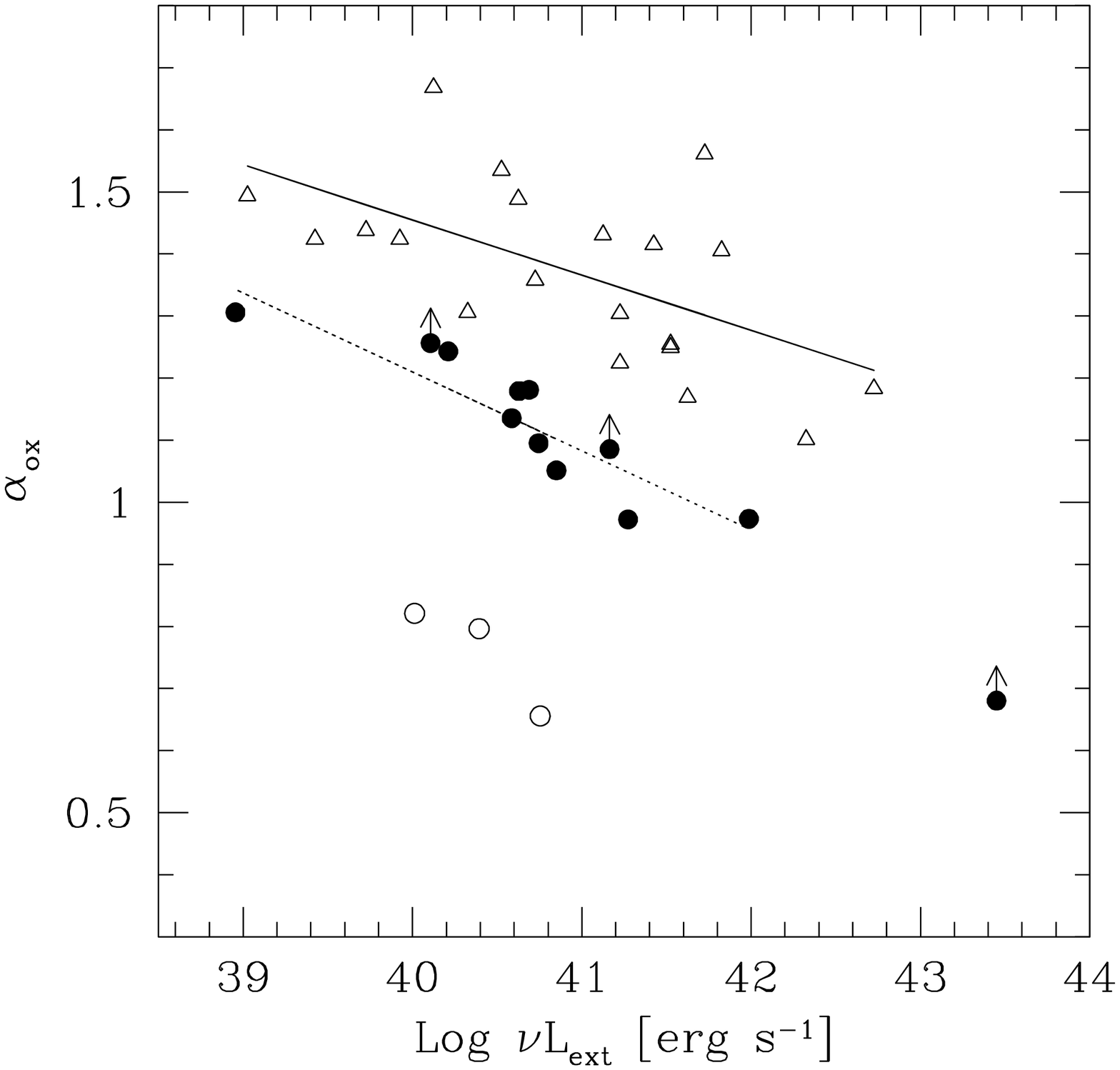,width=0.33\linewidth}}
\caption{Broad-band spectral indices plotted versus the extended radio
luminosity   at  178   MHz   for  FR~I   (filled   circles)  and   LBL
(triangles).  (left) radio/X-ray,  (middle) radio-optical  and (right)
optical/X-ray.  The empty circles represent the FR~I sources for which
we have detected  a large ($N_H > 10^{22}  $cm$^{-2}$) intrinsic X-ray
absorption.  The solid and dotted  lines represent the best linear fit
for the  changes of the spectral  indices with $L_{ext}$  for FR~I and
LBL respectively.}
\label{alpha-rext} 
\end{figure*}

A more detailed comparison  with LBL is particularly instructive.  The
unified model for FR~I and BL~Lacs predicts that these two classes are
drawn from the  same population and are just  seen at different angles
with respect to the jet axis. In Fig.\,\ref{alpha-rext} we compare the
spectral  indices of  the  two  groups taking  also  into account  the
extended  radio-luminosity  L$_{ext}$.  Being  an  isotropic  quantity
L$_{ext}$ does not depend on orientation and indeed LBL and FR~I share
the same range of L$_{ext}$. Thus FR~I overlap in the spectral indices
plane  with  the  jet-dominated  class  with  the  'proper'  level  of
radio-luminosity.

The  differences between  FR~I and  LBL, albeit  quite  small, warrant
further attention  to this point. In  fact, FR~I appear  to be located
only in  the upper left corner  characteristic of the  LBL region (see
Fig.\, \ref{roox}). Fig.\,\ref{alpha-rext} shows  that this is not due
to a  mismatch of extended  radio-luminosity between the  two classes.
However,    as    already     discussed    by    \citet{marco3}    and
\citet{trussoni03},  Doppler  beaming  not  only affects  the  angular
pattern of the  jet emission, but it also causes  a shift in frequency
which in turn  modifies the observed spectral indices.   It is clearly
important to  assess if the  observed (small) differences  between the
spectral indices are consistent with this idea.

In  Fig. \,  \ref{evol}, top  panel, we  give a  simplified  sketch to
graphically illustrate  the origin  of this effect;  we adopted  a SED
characteristic of a LBL (solid line), with the synchrotron peak in the
FIR-mm region and  a flat X-ray spectrum.  The  FR~I SED (dotted line,
re-normalized for  clarity to have  the same level of  the synchrotron
peak as the  BL~Lac) is shifted at lower energies  and this produces a
substantial  decrease   (relative  to   the  radio)  in   the  optical
emission.  Conversely the X-ray  emission is  (always relative  to the
radio) substantially unaltered.  Indeed, no significant differences in
$\alpha_{rx}$ between FR~I and  LBL are present.  Given the relatively
large errors  in the quantities  involved in this comparison,  and the
possible effects  of optical absorption described above,  we prefer to
refrain from a quantitative analysis. Nonetheless, this indicates that
the small differences between the spectral indices of FR~I and BL~Lacs
can be qualitatively accounted for with the presence of beaming.

\subsection{Spectral indices evolution with luminosity}

Fig.\,\ref{alpha-rext} provides  a further element in favour  of a jet
origin for the X-ray cores in  FR~I.  Considering both FR~I and LBL, a
clear trend  of the  $\alpha_{ro}$ and $\alpha_{ox}$  spectral indices
with radio  luminosity is present:  more powerful sources  have higher
$\alpha_{ro}$ (and lower $\alpha_{ox}$)  with respect to less luminous
objects,  while  no  trend  between  $\alpha_{rx}$  and  $L_{ext}$  is
apparent.  FR~I thus  appear to follow the same  evolution of spectral
indices with the extended radio luminosity of LBL.

The  slopes of  the correlations  between radio/optical  and optical/X
nuclear luminosity  provide a different  view of the same  effect.  In
Section \ref{corr-sec}  we reported that the dependence  of $L{_o}$ on
$L_{r}$ is  flatter than unity (m=$0.82\pm0.11$), an  indication of an
evolution with  luminosity of  the radio/optical ratio.   The opposite
behaviour is displayed by the $L_{X}$ / $L_{o}$ relationship, which is
steeper than unity (m=$1.16\pm0.09$).

\begin{figure}
\centerline{ \psfig{figure=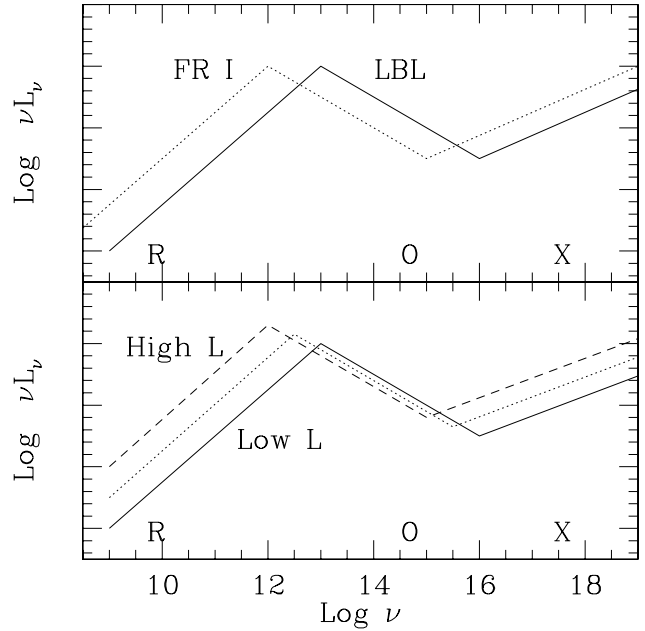,width=1.0\linewidth}}
\caption{Sketch of the SED  evolution related to (top) frequency shift
due to relativistic beaming and (bottom) a change in luminosity .  See
text for details.}
\label{evol} 
\end{figure}

A dependence of the SED on luminosity, over the whole class of blazar,
was recognized by  \citet{fossati98} and it is usually  referred to as
the ``blazar  sequence''.  \cite{ghisellini98} suggested  that this is
due to a relationship between the positions of the emission peaks with
luminosity,  related   to  the  increasing  cooling   effects  on  the
relativistic electrons.  In particular, the synchrotron peak shifts at
lower frequencies in the most powerful sources.  Here we find evidence
for a similar sequence among LBL, although it must be pointed out that
we here consider the extended radio luminosity while \citet{fossati98}
used the radio-core luminosity.  The optical emission, which in LBL is
associated to the decreasing  tail of the synchrotron component, would
be increasingly reduced with the  respect to the radio when the source
luminosity increases (see Fig.\ref{evol} for a sketch).  This accounts
for  the trend observed  in Fig.\,\ref{alpha-rext}  in LBL.  A similar
effect is  seen also in  FR~I and suggests  the presence of  a similar
luminosity  sequence.  This  adds  further weight  to  the  similarity
between  BL~Lacs and  FR~I  and to  a  jet origin  for  the nuclei  in
radio-galaxies.
 
\section{Summary and conclusions}
\label{summary}

We presented  results from Chandra observations of  the 3C/FR~I sample
of low luminosity radio-galaxies. Data are available for 18 out of the
33  sources  of  the  sample  and  we  detected  a  power-law  nuclear
component,  the  characteristic  signature  of  AGN  emission,  in  12
objects.

In 4 galaxies we detected nuclear  X-ray absorption at a level of $N_H
\sim (0.2-6)  \times 10^{22}$  cm$^{-2}$.  X-ray absorbed  sources are
associated  with the presence  of either  highly inclined  dusty disks
seen in  the HST images or  to dust filaments  directly projected onto
the nuclei.  Dust is seen also in most of the remaining sources but no
X-ray absorption  is found  with upper limits  in the range  $N_H \leq
10^{21} -  10^{22}$ cm$^{-2}$.   This suggests an  association between
dust (optical) and  gas (X-ray) absorption but only  when the geometry
is  favourable,  supporting the  existence  of  a flattened  absorber,
reminiscent of  the standard thick  tori envisaged by the  AGN unified
models.  However,  both the X-ray  and (the indirect estimate  of) the
optical absorption, are substantially smaller when compared to what is
typical  of e.g.  Seyfert galaxies.   This indicates  that we  have an
unobstructed view  toward most FR~I  nuclei and that  absorption plays
only a marginal role in the remaining objects.

Concerning the  properties of the obscuring medium,  the ratio between
optical and X-ray  absorption are in the range $A_V /  N_H \sim (0.7 -
3.3) \times 10^{-22}$  mag cm$^2$, slightly smaller  than the standard
galactic value, but not as extreme as found in other classes of AGN.

The  most important issue  is clearly  the origin  of the  X-ray cores
which  can be  produced  either by  the  accretion process  or by  the
associated outflow.   Three results  support the interpretation  for a
common  origin for  the radio,  optical and  X-ray cores,  most likely
associated  to the  base of  a relativistic  jet: i)  the  presence of
strong  correlations  between the  luminosities  in  the three  bands,
extending  over  4  orders  of  magnitude  and  with  a  much  smaller
dispersion ($\sim$ 0.3 dex) when  compared to similar trends found for
other  classes of  AGNs; ii)  the close  similarity of  the broad-band
spectral  indices with the  sub-class of  BL Lacs  (for which  the jet
origin  is  well  established)  sharing  the same  range  of  extended
radio-luminosity, in accord with  the FR~I/BL~Lacs unified model; iii)
the presence of a common evolution of spectral indices with luminosity
in  both  FR~I  and  BL~Lacs,  most  likely due  to  a  shift  in  the
synchrotron emission peak.

In the framework of a jet  origin for the optical and X-ray cores, our
analysis  sets  limits to  the  emission  related  to accretion  at  a
fraction as small as $\sim 10^{-7}$ of the Eddington luminosity, for a
$10^9 M_{\sun}$  black hole.   Thus, regardless of  the origin  of the
optical and  X-ray nuclei,  the luminosity of  LLRG in the  X-ray band
provides further  evidence for a low  level of accretion  and/or for a
low radiative  efficiency of the accretion process.   This was already
noted  by  \citet{chiaberge:ccc}  from  the analysis  of  the  optical
images.   The newly  measured  X-ray luminosities  point  to the  same
conclusion  with the  substantial  advantage of  being  free from  the
lingering  doubts related  to i)  obscuration for  the  optical cores,
since the  X-ray spectra  can be directly  modeled accounting  for the
effects of absorption and ii) the possibility that, due to a different
SED  behaviour with  respect  to  brighter AGN,  most  of the  nuclear
emission is shifted at higher  energies than the optical, as predicted
by        models       of        low        efficiency       accretion
\citep[e.g.][]{narayan00,dimatteo00}.

\acknowledgements  We thank  Marco Chiaberge  for useful  comments and
discussions during the preparation  of the manuscript and the referee,
Martin  Hardcastle, for  his comments  and suggestions  which improved
this paper.  This work was  partly supported by the Italian MIUR under
grant Cofin 2003/2003027534\_002.

\end{document}